# Neural Network-Guided Symbolic Regression for Interpretable Descriptor Discovery in Perovskite Catalysts

Yeming Xian, Xiaoming Wang, Yanfa Yan[*]

*Department of Physics & Astronomy, and Wright Center for Photovoltaics Innovation and Commercialization, The University of Toledo, Toledo, OH, 43606, USA*

*Corresponding author: yanfa.yan@utoledo.edu

**Abstract**

Understanding and predicting the activity of oxide perovskite catalysts for the oxygen evolution reaction (OER) requires descriptors that are both accurate and physically interpretable. While symbolic regression (SR) offers a path to discover such formulas, its performance degrades with high-dimensional inputs and small datasets. We present a two-phase framework that combines neural networks (NN), feature importance analysis, and symbolic regression (SR) to discover interpretable descriptors for OER activity in oxide perovskites. In Phase I, using a small dataset and seven structural features, we reproduce and improve the known μ/t descriptor by engineering composite features and applying symbolic regression, achieving training and validation MAEs of 22.8 and 20.8 meV, respectively. In Phase II, we expand to 164 features, reduce dimensionality, and identify LUMO energy as a key electronic descriptor. A final formula using μ/t, μ/$R_A$, and LUMO energy achieves improved accuracy (training and validation MAEs of 22.1 and 20.6 meV) with strong physical interpretability. Our results demonstrate that NN-guided symbolic regression enables accurate, interpretable, and physically meaningful descriptor discovery in data-scarce regimes, indicating interpretability need not sacrifice accuracy for materials informatics.

## Introduction

The design of efficient electrocatalysts for the oxygen evolution reaction (OER) is central to improving renewable energy technologies such as water splitting and metal-air batteries.[1–4] Oxide perovskites ($ABO_3$) have emerged as promising catalysts due to their structural tunability and earth-abundant components.[5–8] However, experimental synthesis and testing of perovskites remain time-consuming and resource-intensive, motivating the development of predictive models that can identify high-performance candidates from compositional features alone.[9–11]

Recent efforts have demonstrated that machine learning (ML) can accelerate catalyst discovery by learning structure-property relationships from available data.[12–17] Among these, symbolic regression (SR) offers a unique advantage: instead of relying on black-box predictions, it produces human-interpretable equations that link input descriptors to catalytic activity.[18–20] Notably, Weng et al.[18] used symbolic regression to identify a physically meaningful descriptor, μ/t (the ratio of octahedral and tolerance factors), for predicting the OER activity of oxide perovskites. While promising, this approach has key limitations: SR methods like *gplearn* struggle in high-dimensional feature spaces, especially when training data are scarce. As Figure S1 and Table S1 show, we tested the ability of *gplearn* to generate interpretable analytical formula with the same dataset. Although the prediction accuracy for the training set increases with increased formula complexity, the formulas are less interpretable as most features are included, with only a subset being highly predictive. Without embedded feature selection, symbolic regressors often explore irrelevant variables, leading to unstable models or overfitting.

To address this, we proposed a hybrid workflow that combines neural networks (NNs) with symbolic regression for interpretable catalyst modeling. Neural networks were first used to learn from the data and identify important features via permutation importance analysis.[11,15,21] Only the most predictive features were then passed to the symbolic regressor (*gplearn*), which searches for concise formulas. This strategy improves both the stability and interpretability of the final descriptors. Compared to using SR alone, our NN-guided pipeline avoids irrelevant variables and reduces the search space to meaningful feature combinations.

We benchmarked our method against conventional regressors—including Ridge, random forest, XGBoost, and Principal Component Analysis (PCA)-augmented models—all of which underperform in this small-data, high-dimensional regime.[13,16,21] Our results showed that combining neural networks for feature ranking with symbolic regression enables accurate and interpretable descriptor discovery, offering a reproducible framework for data-scarce materials problems.

# Results

We adopted a two-phase strategy (Figure 1) to develop an interpretable model. In Phase I, we reproduced and improved earlier symbolic regression results by combining neural networks with permutation importance. From the original seven features, we identified μ, $R_A$, and t as key variables, engineered composite features (μ/t, μ/$R_A$, $R_A$·t), and used symbolic regression to derive a compact formula with improved accuracy. In Phase II, we expanded the feature space using six *Matminer* featurizers and applied filtering to reduce dimensionality. Neural network analysis identified LUMO energy as a new, physically meaningful descriptor. Using μ/t, μ/$R_A$, and LUMO, symbolic regression produced an interpretable formula with better validation MAE. This workflow yields concise formulas with improved predictive accuracy without sacrificing the interpretability.

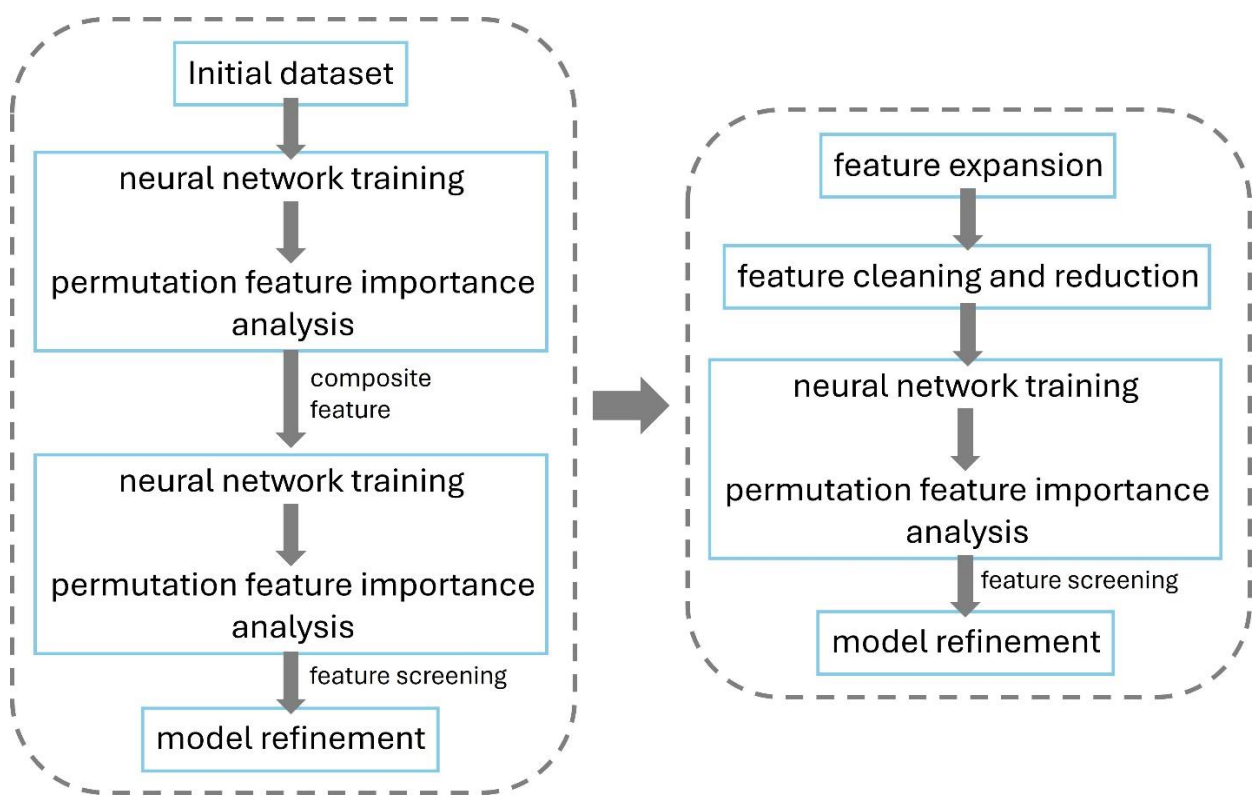

Figure 1. Workflow of the two-phase modeling framework. Phase I uses the original small dataset; Phase II expands the feature set. Both phases involve neural-network modeling, permutation-based importance analysis, and symbolic regression via *gplearn* to derive analytic formulas.

**Phase I – Initial Dataset Analysis**

**Key Features Identified**

In Phase I, we trained a neural network model using the original seven physical descriptors to predict OER activity, as represented by the overpotential vs. the reversible hydrogen electrode ($V_{RHE}$), across the 18 training and 5 validation samples. The model architecture consists of three hidden layers with 128, 64, and 16 neurons respectively, using ReLU activation and L2 regularization, followed by a linear output layer for regression (Figure 2a). During training, the model demonstrated stable convergence, achieving a final training MAE of 0.0236 eV and a validation MAE of 0.0243 eV, with consistent decreases in both loss and MAE over epochs (Figure S2).

To interpret the model, we computed permutation feature importance scores. This analysis revealed that the octahedral factor ($\mu$), the tolerance factor (t), and the A-site ionic radius ($R_A$) are the most influential features (Figure 2c). These three features consistently contributed most to prediction accuracy, suggesting that OER activity is primarily governed by the structural geometry of the perovskite lattice. The prediction–experiment correlation plot (Figure 2b) further confirms the model's reliability in capturing underlying trends in the data.

To better understand the physical role of the top features, we examined their individual relationships with the measured $V_{RHE}$. As shown in Figure S3, $V_{RHE}$ varies monotonically with respect to $\mu$, t, and $R_A$, confirming their direct relevance to catalytic activity. These findings provided the basis for generating composite features in the next step, including $\mu/t$, $\mu/R_A$, and $R_A \cdot t$, which are used for model refinement and symbolic regression in the subsequent analysis.

**Feature Engineering and Retraining**

Building on the insights from the initial neural network analysis, we constructed three composite features—$\mu/t$, $\mu/R_A$, and $R_A \cdot t$—to capture physically meaningful interactions among the most important variables. These combinations reflect known geometric and structural relationships in perovskite materials and were added to the input feature set for retraining the neural network. After retraining with the expanded feature set, the neural network showed improved predictive performance, achieving a reduced training MAE of 0.0231 eV and validation MAE of 0.0210 eV, as shown in the updated training metrics (Figure S4). This performance improvement demonstrates that the engineered features provided additional predictive power beyond the original inputs.

The updated model's predictions again showed strong agreement with measured $V_{RHE}$ values (Figure 3a), confirming that the network captured the relevant trends. Permutation feature importance analysis on the retrained model (Figure 3b) indicated that the newly added composite

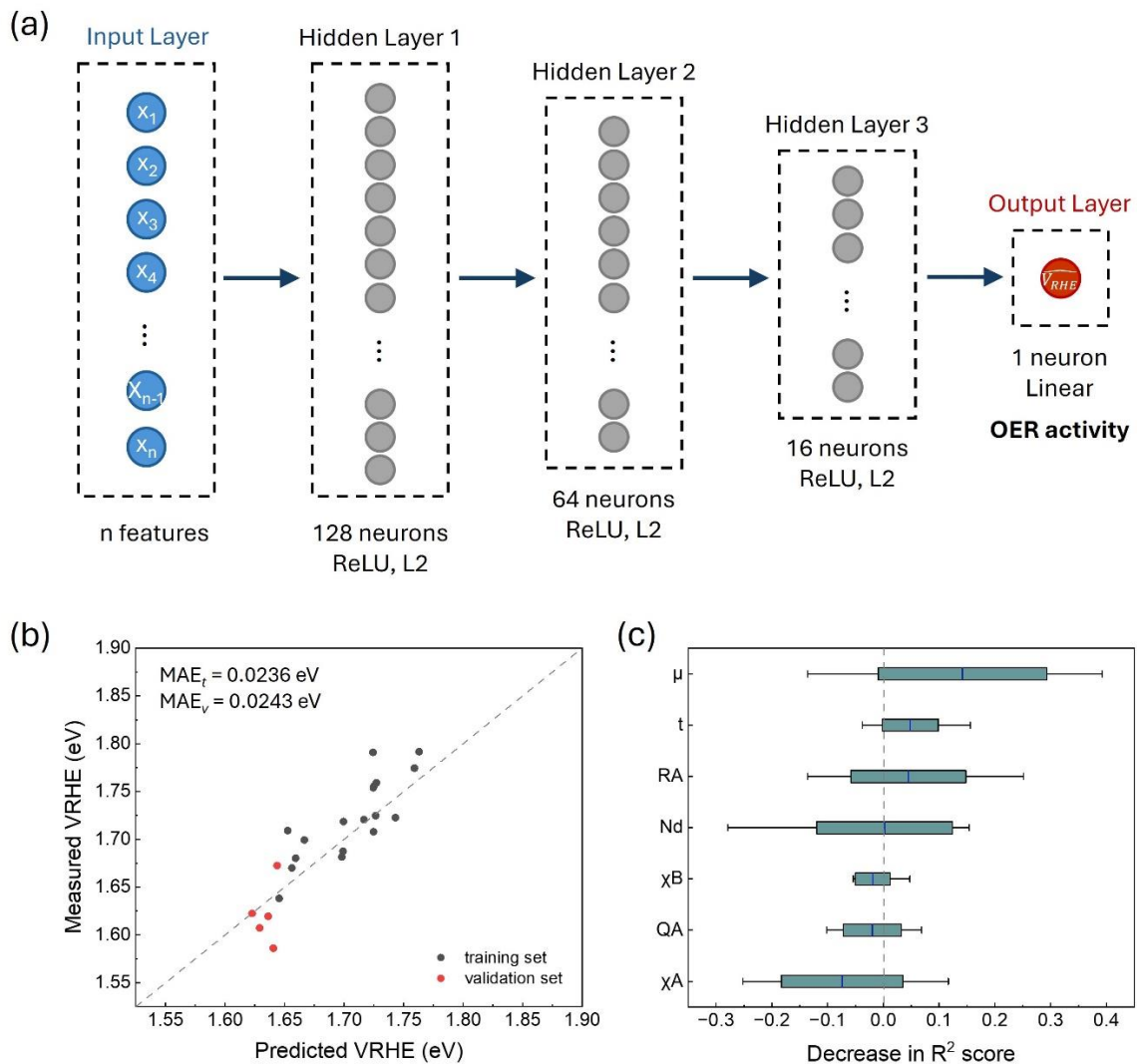

Figure 2. (a) Neural network architecture used for Phase I/II modeling. The network consists of three hidden layers with 128, 64, and 16 neurons respectively (ReLU activation and L2 regularization), and a linear output layer predicting OER activity ($V_{RHE}$). This architecture was used to identify important features and guide symbolic regression. (b) Neural network performance. Predicted versus measured $V_{RHE}$ values for the training and validation set, showing strong agreement and low prediction error. (c) Feature importance in Phase I (initial feature set). Permutation feature importance ranking, indicating that μ, t, and $R_A$ are the most influential descriptors in the model.

features—especially μ/t and μ/$R_A$—were among the most influential inputs, validating the effectiveness of this feature engineering step.

To further support their physical relevance, we analyzed the dependence of experimental $V_{RHE}$ on the three composite descriptors (Figure S5). Each showed a clear and interpretable relationship with catalytic activity, reinforcing their suitability as candidate inputs for symbolic regression. These composite features were therefore selected as the basis for symbolic formula discovery in the next stage of the workflow.

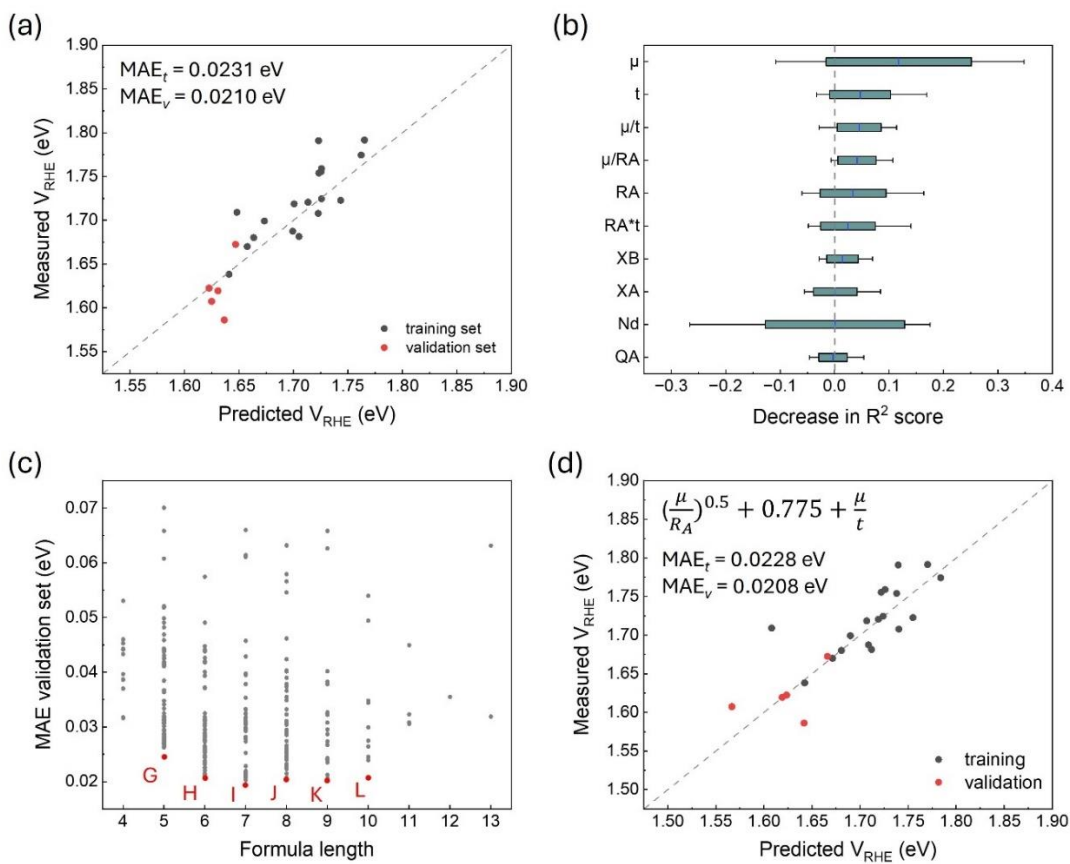

Figure 3. (a) Predicted versus measured $V_{RHE}$ values using the neural network model retrained with original and composite features (μ/t, μ/$R_A$, $R_A$·t). (b) Permutation feature importance scores for the retrained model. The composite features μ/t and μ/$R_A$ rank among the most important, confirming their predictive value. (c) Validation MAE of symbolic regression formulas generated by *gplearn*, plotted against formula length. Each point represents one of 432 formulas produced under different hyperparameter settings. (d) Performance of the symbolic regression formula corresponding to Point H on the training and validation sets.

**Symbolic Regression Formula**

Following neural network training and feature importance analysis, we applied symbolic regression using the gplearn package to extract interpretable analytical formulas linking descriptors to $V_{RHE}$. The symbolic regressor was provided with 6 refined features t, μ, $R_A$, μ/t, μ/$R_A$, and $R_A·t$. To ensure thorough coverage of the model space, we performed a grid search over 432 combinations of symbolic regression hyperparameters, resulting in 432 unique formulas.

Figure 3c and S6 show the distribution of mean absolute error (MAE) across all generated formulas for the validation and training sets, plotted against formula length. Each point represents a candidate formula, and the markers (A–L) highlight the best-performing formula in each formula-length group. As expected, shorter formulas generally yielded higher MAEs, while some expressions achieved lower error but at the cost of complexity. Among these, the twelve representative formulas (A-L) were selected for further analysis, as summarized in Table S2. We evaluated each in terms of accuracy and interpretability. Four formulas corresponding to A, G, H, and L—were considered physically rational, meaning they exhibited correct monotonic trends between $V_{RHE}$ and the involved features as the original dataset renders. Notably, formulas H and L achieved the best performance, with training MAEs of 0.0228 eV and 0.0229 eV, and validation MAEs of 0.0208 eV and 0.0209 eV, respectively.

As shown in Figure 3d and S7, both formulas H and L produced predictions that closely match the measured $V_{RHE}$ values on the training and validation set, confirming their generalizability. Moreover, both expressions involved the composite features μ/t and μ/$R_A$, supporting the results of the feature engineering stage. These formulas balance simplicity, accuracy, and physical interpretability, making them strong candidates for use as predictive descriptors in perovskite catalyst design.

**Phase II – Expanded Feature Space**

**New Features and Importance**

In Phase II, we expanded the input feature space using the *Matminer* library to capture additional chemical, electronic, and structural descriptors relevant to OER activity. As shown in Figure 4a, six *Matminer* featurizers—ElementProperty, Stoichiometry, ValenceOrbital, OxidationStates, AtomicOrbitals, and BandCenter—were used to generate 164 numerical features. After removing near-constant features via a variance filter (threshold = $10^{-4}$), 103 features remained. We further applied a Pearson correlation matrix for this intermediate feature set (Figure S8), which illustrates the degree of redundancy present before filtering, to eliminate highly correlated pairs (coefficient threshold = 0.95), resulting in 46 uncorrelated features. Importantly, the original seven features and the three engineered composite features from Phase I were always retained during this process. The final feature set thus combined expert-informed and data-driven descriptors.

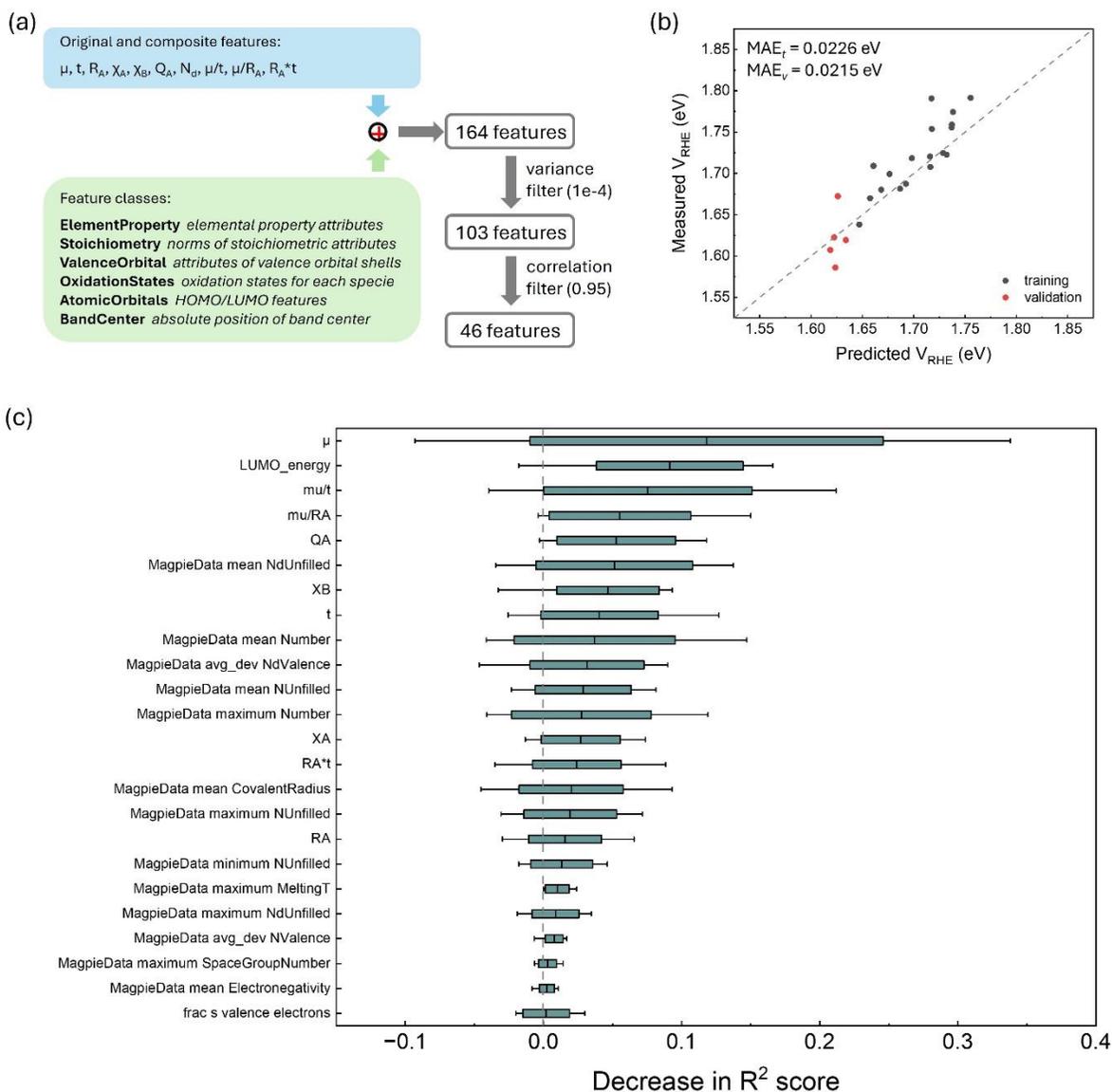

Figure 4. (a) Feature expansion and reduction workflow. *Matminer* generated 164 numerical features, which were reduced to 46 using variance and Pearson correlation filters. Original and composite features were retained throughout. (b) Predicted versus measured $V_{RHE}$ using the neural network trained on the reduced 46-feature dataset. (c) Permutation feature importance for the Phase II neural network model. Among the newly added features, LUMO energy emerged as one of the most important descriptors, alongside previously engineered structural features such as μ/t and μ/$R_A$.

The filtered 46-feature dataset was then used to retrain the neural network in Phase II. As shown in Figure S9, the model achieved strong performance, with a training MAE of 0.0226 eV and validation MAE of 0.0215 eV, indicating good generalization despite the expanded feature space. The prediction–measurement correlation plot (Figure 4b) shows a tight clustering along the diagonal, further supporting the reliability of the NN model.

To benchmark the neural network against traditional ML models under the same conditions (23 samples, 46 features), we tested five additional frameworks: Ridge regression, random forest (RF), XGBoost, XGBoost with DART booster, and PCA + XGBoost. Figure S10-S16 summarize hyperparameter tuning and performance comparisons for each traditional model. The Ridge model (Figure S10) showed limited generalization: while regularization strength influenced training behavior, validation MAE remained significantly higher than the NN. Tree-based models including RF and XGBoost (Figure S11-S15) performed well on training data but overfit the small dataset, with large validation MAEs. Even after applying PCA to reduce dimensionality to nine principal components (Figure S16), XGBoost did not outperform the NN. The model performance comparison (Table 1) shows that the neural network consistently achieves the lowest validation MAE compared to traditional methods. These results underscore a key limitation of traditional ML models in small, high-dimensional datasets: Ridge regression lacks the representational flexibility to capture nonlinear relationships, while ensemble trees tend to overfit despite extensive hyperparameter tuning. In contrast, the use of weight regularization in NN provides a more stable balance between model complexity and generalization.

Permutation feature importance analysis (Figure 4c) revealed a new electronic descriptor—LUMO energy—as one of the most influential features in the model. The LUMO feature originates from *Matminer*'s AtomicOrbitals featurizers and reflects the energy of the lowest unoccupied molecular orbital. Its prominence in the NN model suggests a strong link between electronic structure and catalytic performance, and supports its inclusion in the symbolic regression step to follow.

Table 1. Validation MAE comparison across six ML models trained on the (23, 46) dataset.

| Model | Validation MAE (meV) |
|---|---|
| Neural Network | 21.5 |
| Ridge Regression | 44.1 |
| Random Forest | 52.2 |
| XGBoost | 36.0 |
| XGBoost with DART | 37.9 |
| PCA + XGBoost | 60.7 |

**Physical Interpretability of LUMO**

A key advantage of our neural-symbolic framework is the ability to identify not only accurate but also physically meaningful descriptors. As discussed above, among the new features introduced in Phase II, LUMO energy emerged as a top-ranked variable from the permutation feature importance analysis. In *Matminer*, the LUMO energy of a compound is estimated using the AtomicOrbitals featurizer, which calculates a composition-weighted average of atomic LUMO levels. Each atomic LUMO is empirically derived from electronegativity and valence orbital type—more electronegative atoms tend to have deeper (more negative) LUMO energies. To understand its relevance, we examined the relationship between measured $V_{RHE}$ and LUMO energy across all 23 perovskite samples. As shown in Figure 5a, the data exhibit a non-monotonic trend: the lowest $V_{RHE}$ values occur when the LUMO energy lies in an optimal range between approximately -0.34 eV and -0.32 eV. Outside this range, VRHE increases—suggesting that either too shallow or too deep a LUMO reduces catalytic efficiency. This implies that LUMO energy influences OER activity in a more nuanced way than purely monotonic trends.

The energy level of the LUMO relative to the Fermi level critically impacts OER activity by influencing the binding strength of reaction intermediates (e.g., OH*, OOH*) on perovskite surface, as Figure 5b shows. A LUMO energy that is too high, significantly above the Fermi level, results in weak interactions with the highest occupied molecular orbitals (HOMOs) of OER intermediates, leading to insufficient adsorption and high energy barriers for initiating reaction steps, thus reducing OER activity and requiring higher overpotentials ($V_{RHE}$) to drive the reaction. Conversely, a LUMO energy that is too low, overly close to or below the Fermi level, causes excessively strong binding of intermediates, hindering their desorption (e.g., $O_2$ release), which also increases overpotential and diminishes catalytic efficiency. According to the Sabatier principle,[22] optimal OER performance occurs when reaction intermediates moderately bind to perovskite surface. Therefore, the LUMO energy that is moderately above the Fermi level balances adsorption and desorption energetics, as seen in catalysts like $Sr_{0.25}Ba_{0.75}NiO_3$ and $Cs_{0.4}La_{0.6}Mn_{0.25}Co_{0.75}O_3$ in our dataset, and enhances overall OER activity.

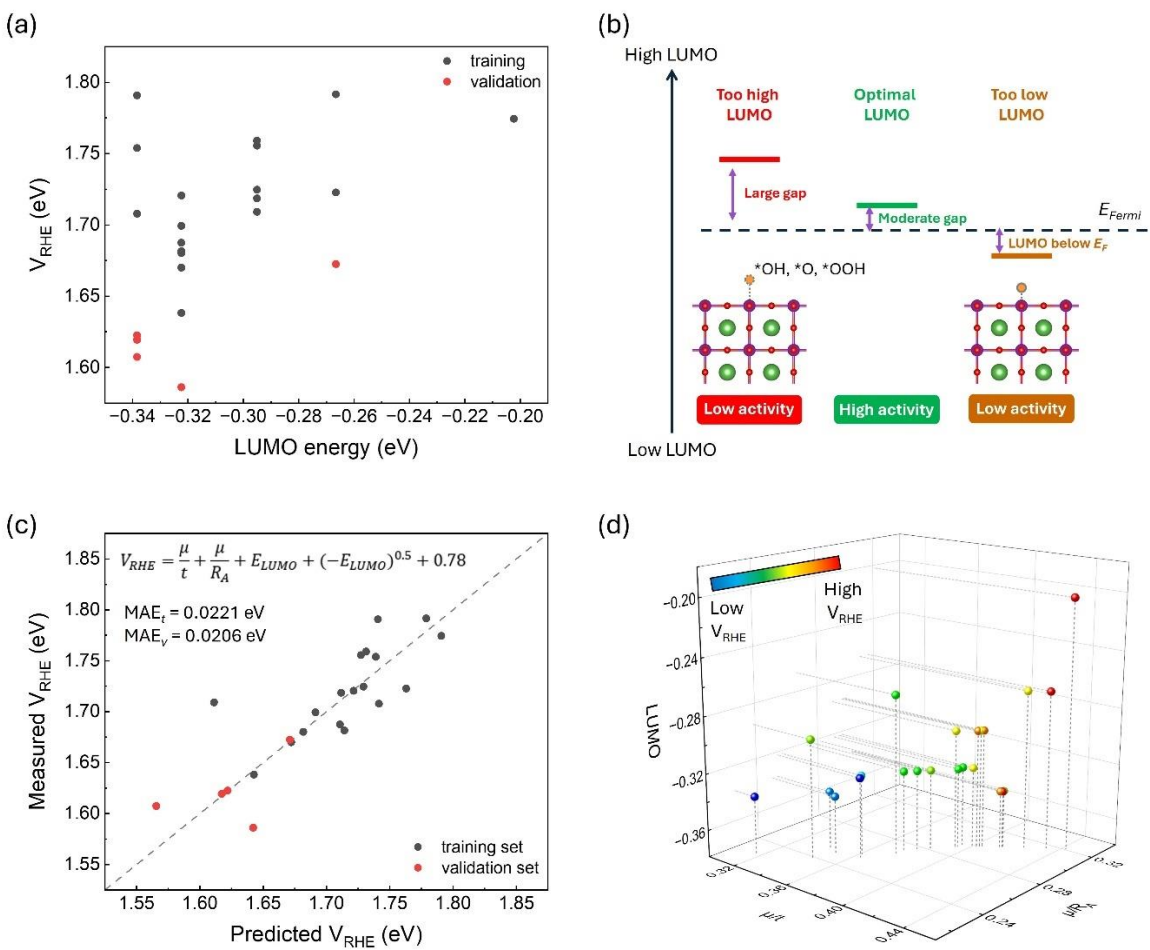

Figure 5. (a) Relationship between LUMO energy and measured $V_{RHE}$ for 23 oxide perovskite samples. (b) Schematic illustration of the variance of OER activity as a result of LUMO level fluctuation with respect to the Fermi energy. (c) Predicted versus measured $V_{RHE}$ using the refined symbolic regression formula involving $\mu/t$, $\mu/R_A$, and LUMO energy. (d) 3D visualization of measured $V_{RHE}$ as a function of $\mu/t$, $\mu/R_A$, and LUMO energy for the 23 samples. Blue points indicate low $V_{RHE}$ and optimal catalytic performance.

**Refined Formula and Performance**

To derive an interpretable expression linking structure and electronic properties to OER activity, we applied symbolic regression using *gplearn*, this time using only the three most physically meaningful features identified from the Phase I and II analysis: $\mu/t$, $\mu/R_A$, and LUMO energy. Among the formulas generated across the symbolic regression hyperparameter space, only two candidates incorporated all three features. Their structures, lengths, and associated MAEs on training and validation sets are summarized in Table S3. The better-performing formula achieved a training MAE of 0.0221 eV and a validation MAE of 0.0206 eV, slightly improving upon the Phase I models while maintaining a compact symbolic form. As shown in Figure 5c, this formula yields predictions that align closely with experimental $V_{RHE}$ values, demonstrating its robustness and generalizability.

To further visualize the overall relationship, we generated a 3D scatter plot of measured $V_{RHE}$ as a function of the three input features (Figure 5d). The plot clearly reveals an optimal region in feature space: combinations with low $\mu/t$ and $\mu/R_A$ (i.e., geometrically favorable structures) and LUMO energies between -0.34 and -0.32 eV yield the lowest $V_{RHE}$ values. In contrast, compounds with similar geometric profiles but LUMO energies outside this window show elevated $V_{RHE}$—such as the single green point in the optimal $\mu/t$ and $\mu/R_A$ region but with a high LUMO. This demonstrates that both geometric and electronic factors must simultaneously align to achieve high catalytic performance.

This 3D plot provides a powerful screening guideline: efficient OER catalysts should not only exhibit favorable geometric ratios (low $\mu/t$ and $\mu/R_A$) but also possess LUMO energies in a specific range. The refined formula derived here thus serves as a physically interpretable model for identifying promising oxide perovskite catalysts in future design efforts.

## Discussion

**Interpretation of Descriptors.** The descriptors µ/t, µ/R$_A$, and LUMO energy critically influence the OER performance of perovskites by governing structural stability of perovskites and electronic interactions on perovskite surface. The ratio µ/t, combining the octahedral factor (µ = R$_B$/R$_X$) and tolerance factor (t = (R$_A$+R$_X$)/ $\sqrt{2}$(R$_B$+R$_X$)), reflects B-site cation fit in BX$_6$ octahedra and lattice stability, with high values inducing octahedral tilting that lowers LUMO energy and strengthens OER intermediate adsorption, potentially increasing V$_{RHE}$ if excessive. Similarly, µ/R$_A$ balances B- and A-site cation sizes, where high values cause lattice strain, lowering LUMO energy and hindering O$_2$ release. As discussed above, the LUMO energy directly controls intermediate binding, with optimal positioning moderately above the Fermi level, minimizing V$_{RHE}$ by balancing adsorption/desorption. Therefore, our final formula is fully analytic and interpretable: it can guide new material designs by suggesting to increase *t*, decrease *µ*, and adjust components to acquire optimal LUMO energy.[23]

**Performance.** Across both phases, our approach yields low MAEs on held-out data. For example, the Phase I formula achieved an MAE of 22.8 meV (training) and 20.8 meV (validation) on V$_{RHE}$, while the Phase II formula achieved slightly lower MAEs of 22.1 meV (training) and 20.6 meV (validation), as listed in Table 2. These errors are comparable to or better than previously reported *gplearn* standalone models, demonstrating that interpretability need not sacrifice accuracy.

Table 2. Comparison of the MAE performance for training (MAE$_t$) and validation (MAE$_v$) sets of the analytical formulas generated in previous work and in Phase I and II of this work.

| Analytical formula | (MAE$_t$, MAE$_v$) (meV) |
|---|---|
| $1.554 \frac{\mu}{t} + 1.092$ [18] | (25.3, /) |
| $(\frac{\mu}{R_A})^{0.5} + \frac{\mu}{t} + 0.775$ (Phase I) | (22.8, 20.8) |
| $\frac{\mu}{t} + \frac{\mu}{R_A} + E_{LUMO} + (-E_{LUMO})^{0.5} + 0.780$ (Phase II) | (22.1, 20.6) |

**Advantages of the Framework.** The two-phase neural-symbolic framework presented in this work offers a robust and interpretable approach to modeling catalyst performance in data-scarce materials systems. By combining neural networks for feature importance analysis with symbolic regression for formula discovery, the method balances predictive accuracy and physical interpretability. It effectively filters out irrelevant variables before regression, mitigating overfitting and improving generalization—especially important given the small sample size and high dimensionality of the expanded feature space. The workflow integrates domain knowledge, such as established structural descriptors ($\mu/t$, $\mu/R_A$), with data-driven insights, including the identification of LUMO energy as a key electronic feature.

In this work, we presented a two-phase, interpretable machine learning framework for predicting OER activity in oxide perovskites using neural networks and symbolic regression with small-data setting. In Phase I, we reproduced and improved the known $\mu/t$ descriptor by introducing composite features ($\mu/R_A$) and guiding symbolic regression with neural feature importance. In Phase II, we expanded the feature set with *Matminer*, reduced dimensionality, and identified LUMO energy as a key electronic descriptor. The final symbolic formula, composed of $\mu/t$, $\mu/R_A$, and LUMO energy, achieves the lowest validation error and reveals an optimal range of these features for catalytic performance. Our findings provide both a quantitative predictive model and qualitative design guidelines for efficient perovskite catalysts. More broadly, this neural network-symbolic regression strategy offers a generalizable framework for interpretable, data-efficient modeling across materials science applications.

## Methods

**Dataset and Features (Phase I):** We use the same perovskite dataset as reported by Weng *et al.*: 23 $ABO_3$ materials (18 conventional and 5 new perovskites) with measured OER activity.[18] For $V_{RHE}$ target values, we select the measurement results at current density of 5 mA cm$^{-2}$. We split these 23 samples with the 18 conventional samples as training set and the 5 new samples as validation set. For each material we select seven physical descriptors: the tolerance factor $t$ and octahedral factor $\mu$, the ionic radius of the A-site $R_A$, electronegativities of A-site and B-site $\chi_A$ and $\chi_B$, valence state of A-site $Q_A$, and number of $d$-electrons on TM B-site $N_d$.

**Feature Expansion (Phase II):** In Phase II, we expanded the feature space using the *Matminer* library to generate a broader and more diverse set of physically motivated descriptors.[24] Specifically, we employed six feature classes from *Matminer*—ElementProperty, Stoichiometry,

ValenceOrbital, OxidationStates, AtomicOrbitals, and BandCenter—which together produced a total of 164 numerical features per sample. To improve model robustness and mitigate overfitting in the data-limited regime, we applied two filtering steps. First, a variance filter removed features with a variance below $10^{-4}$ across the dataset, reducing the feature set to 103 features.[25] Second, a correlation filter eliminated highly redundant features by removing one from each pair with Pearson correlation coefficient greater than 0.95, resulting in a final set of 46 uncorrelated, high-variance features.[26] These 46 features were then used for neural network training and permutation importance analysis in Phase II of our workflow.

**Neural Network Training:** A three-layer fully connected neural network implemented in *TensorFlow* with ReLU activation and L2 regularization was trained using the Adam optimizer with an exponentially decaying learning rate. We used a custom early stopping strategy with a warm-up phase to stabilize training. The loss function of mean squared error (MSE) was selected, and model performance was evaluated using mean absolute error (MAE) on both training and validation sets. We used permutation feature importance from *scikit-learn* to identify the most predictive features.[27] The analysis was performed on the validation set using $R^2$ as the scoring metric, with 30 random shuffles per feature to ensure stability. Features causing the greatest drop in model performance when permuted were considered most important and selected for symbolic regression.

**Symbolic Regression with gplearn:** After identifying the most important features using neural network-based permutation importance, we applied symbolic regression using the *gplearn* library to derive interpretable analytic formulas.[20] To ensure a comprehensive search of the symbolic expression space, we performed a grid search over key genetic programming hyperparameters: crossover probability (*p_crossover*) ranged from 0.5 to 0.925 in increments of 0.025, and the remaining mutation probabilities (*p_subtree_mutation*, *p_hoist_mutation*, and *p_point_mutation*) were adjusted accordingly to maintain a total probability of 1, following the hyperparameter grid reported by Weng *et al*. The parsimony coefficient, which penalizes overly complex expressions, varied from 0.0005 to 0.0015 in increments of 0.0005. For each combination of parameters, a symbolic regressor was trained with a population size of 5000, 20 generations, and a stopping criterion of 0.01 for the mean absolute error (MAE). The function set included basic arithmetic operations and unary functions such as *sqrt* and *log*.

# Acknowledgement

The neural network training were supported by the computational resources sponsored by the Department of Energy's Office of Energy Efficiency and Renewable Energy and located at the National Renewable Energy Laboratory, and the resources of the National Energy Research Scientific Computing Center (NERSC), a US Department of Energy Office of Science User Facility located at Lawrence Berkeley National Laboratory, operated under contract DE-AC02 05CH11231 using NERSC award BES-ERCAP0028897 and BES-ERCAP0032847.

Supplementary Material for

# Neural Network-Guided Symbolic Regression for Interpretable Descriptor Discovery in Perovskite Catalysts

Yeming Xian, Xiaoming Wang, Yanfa Yan[*]

*Department of Physics & Astronomy, and Wright Center for Photovoltaics Innovation and Commercialization, The University of Toledo, Toledo, OH, 43606, USA*

This supplementary material include candidate formulas by *gplean*, metric optimization for NN training, the relation between $V_{RHE}$ and singular predictive features, formula performance, Pearson correlation matrix, and parameter tuning and model performance for traditional ML frameworks.

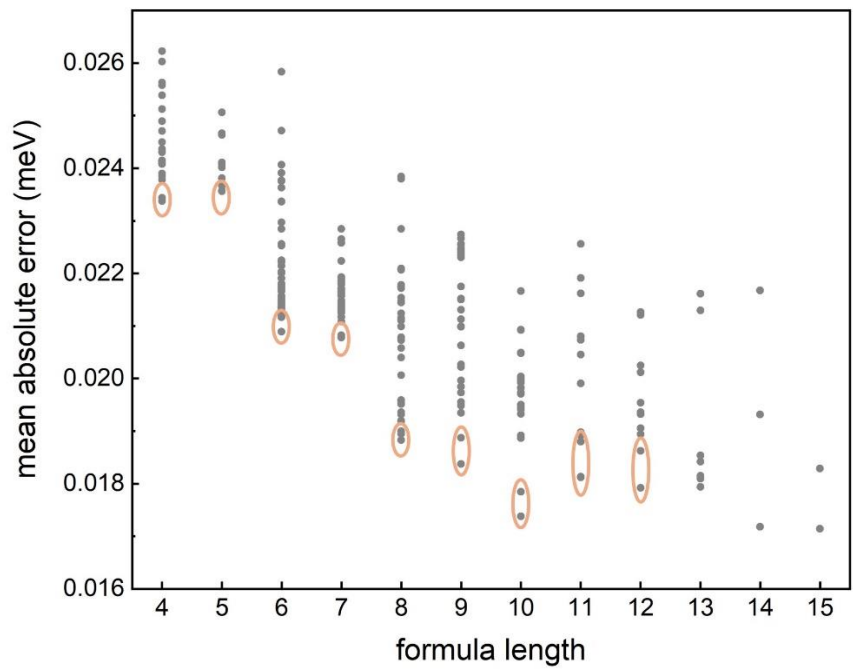

Figure S1. Performance trade-off in symbolic regression without feature filtering. As formula complexity increases, prediction accuracy improves, but interpretability and generalizability degrade.

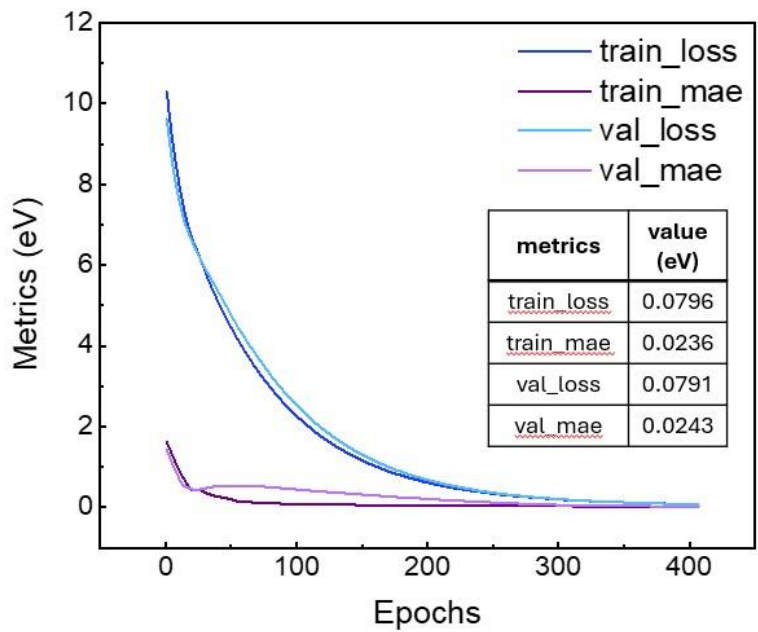

Figure S2. Evolution of training and validation loss and MAE during neural network training with 7 features. The model converges smoothly with good generalization, reaching stable validation performance.

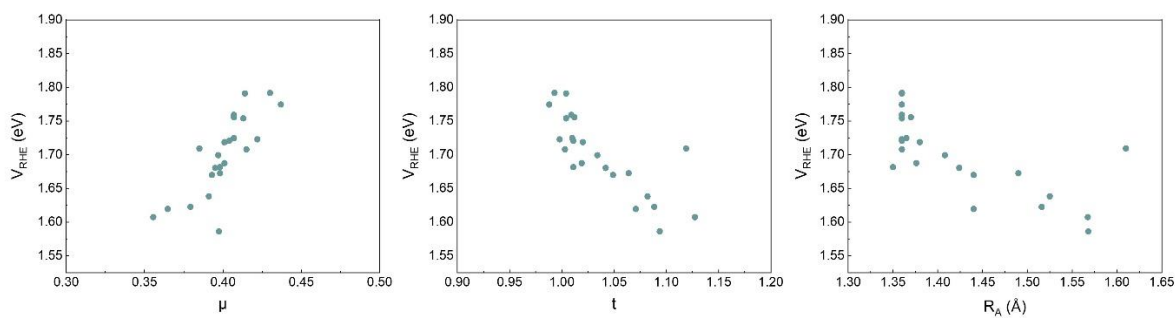

Figure S3. Correlation between measured $V_{RHE}$ and the most important individual features from the initial NN model including $\mu$, $t$, and $R_A$.

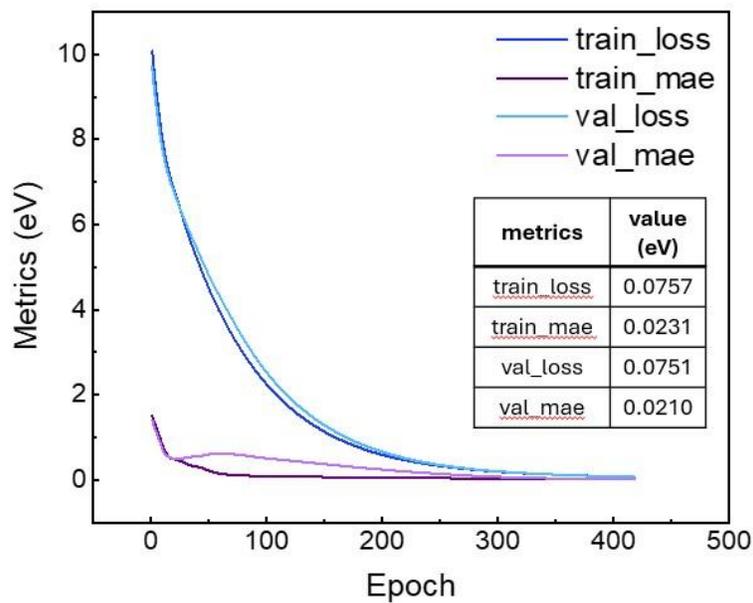

Figure S4. Training evolution after feature engineering, where $\mu/t$, $\mu/R_A$, and $R_A \cdot t$ are added. The network achieves improved MAEs of 0.0231 eV (training) and 0.0210 eV (validation), indicating enhanced predictive performance.

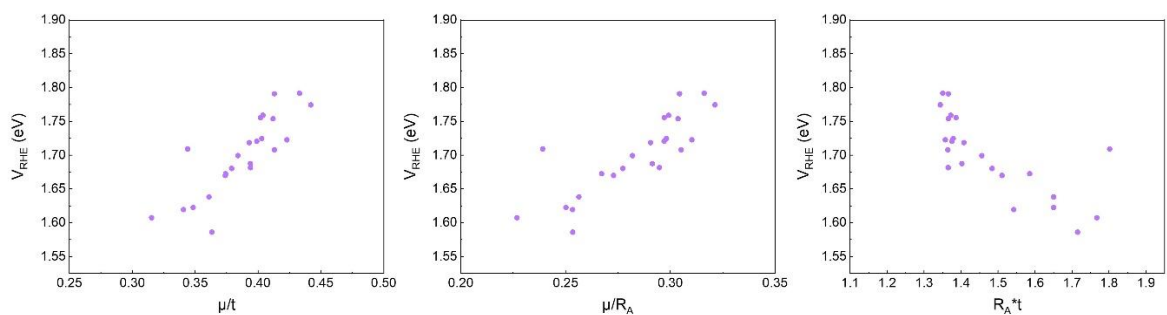

Figure S5. $V_{RHE}$ variation with composite features. Clear monotonic trends confirm physical relevance of $\mu/t$, $\mu/R_A$, and $R_A \cdot t$ in describing OER activity.

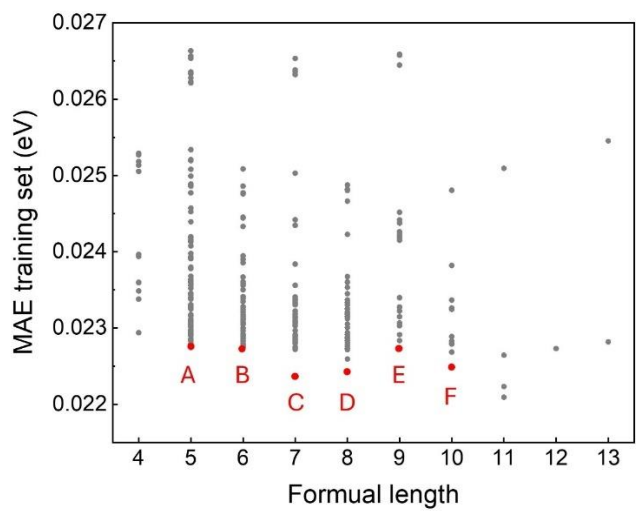

Figure S6. Distribution of MAEs from symbolic regression with composite features. Each dot represents one of 432 symbolic regression formulas. Lower complexity models generally show higher MAEs, while optimal trade-offs lie in medium-length expressions.

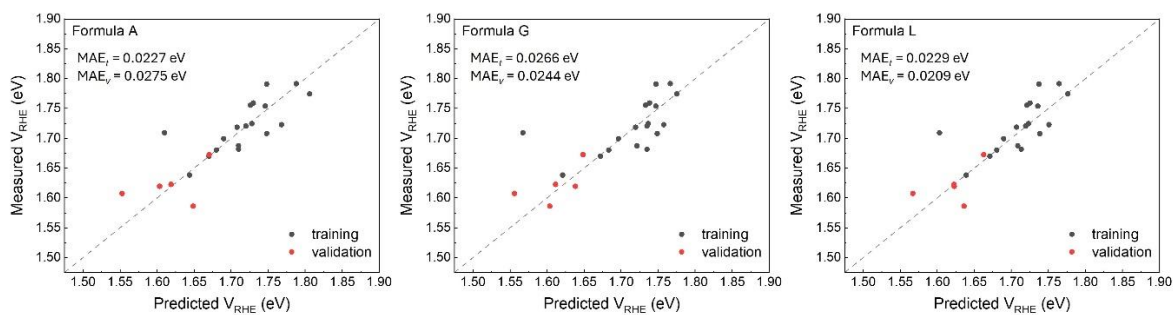

Figure S7. Performance of formulas A, G, and L. Predicted vs. measured $V_{RHE}$ values demonstrate generalization.

Figure S8. Pearson correlation matrix of the 103 features remaining after variance filtering. Strongly correlated pairs were removed based on a Pearson coefficient threshold of 0.95, yielding a final reduced feature set.

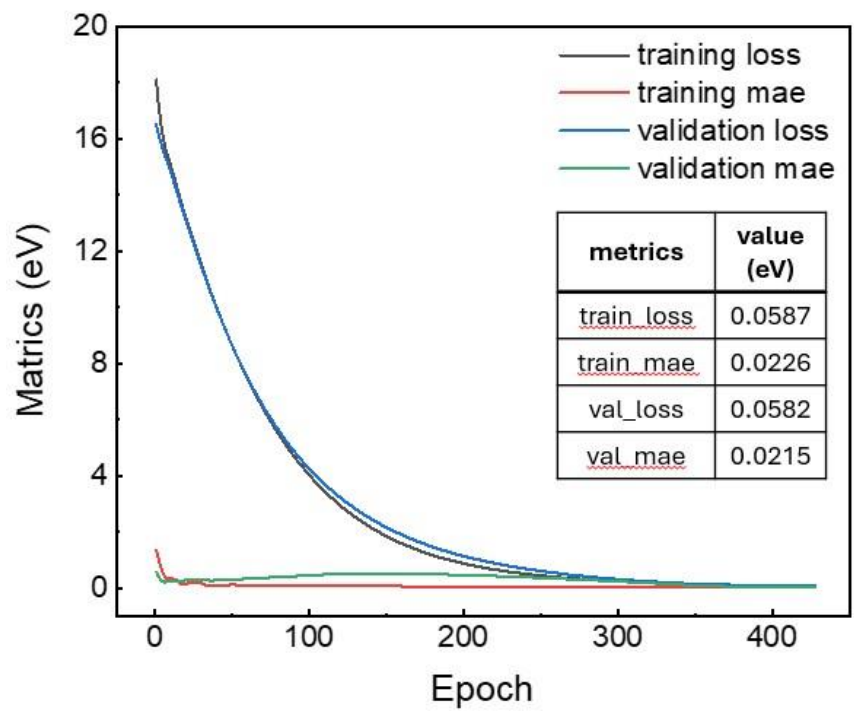

Figure S9. Evolution of training and validation loss and MAE during neural network training on the expanded (23, 46) dataset. The model reaches a training MAE of 0.0226 eV and validation MAE of 0.0215 eV, demonstrating strong generalization despite increased dimensionality.

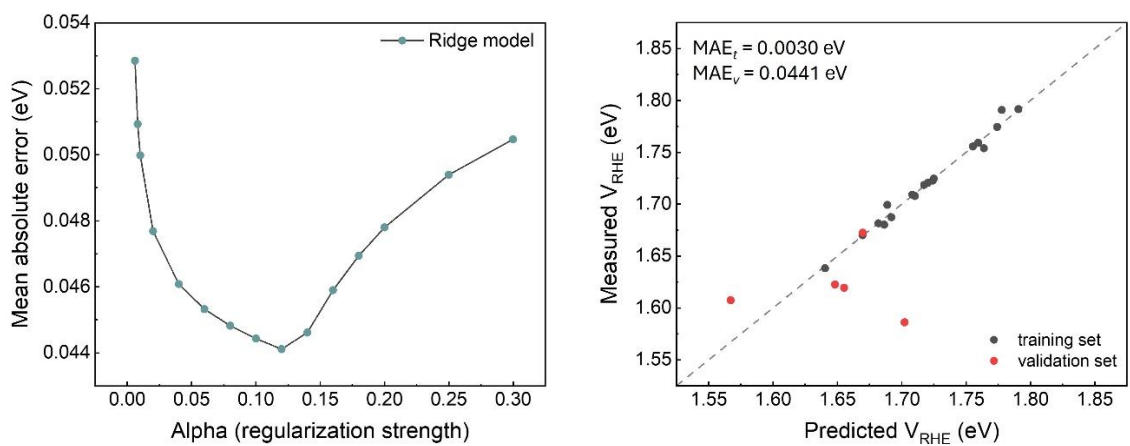

Figure S10. Performance of Ridge regression across different regularization strengths (α). While training MAEs are low, the model shows high validation error due to limited flexibility for capturing nonlinear relationships.

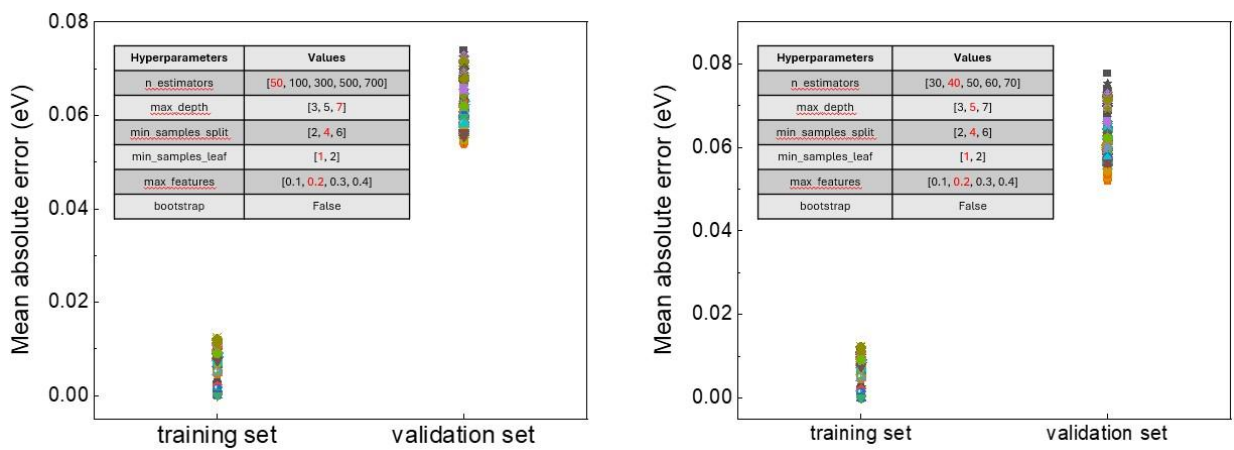

Figure S11. Hyperparameter tuning for the random forest model. Despite extensive optimization, the model overfits the training data, achieving low training MAE but poor validation performance.

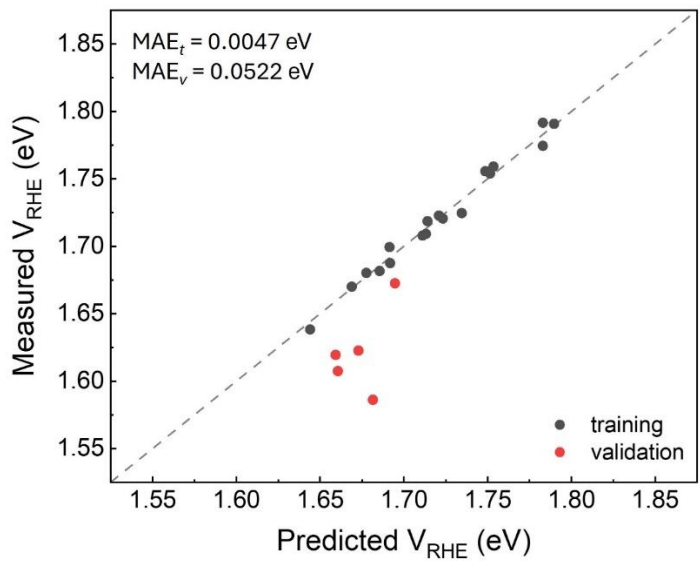

Figure S12. Best random forest model performance. Predicted vs. measured VRHE plot confirms strong overfitting, with low training MAE but large validation MAE.

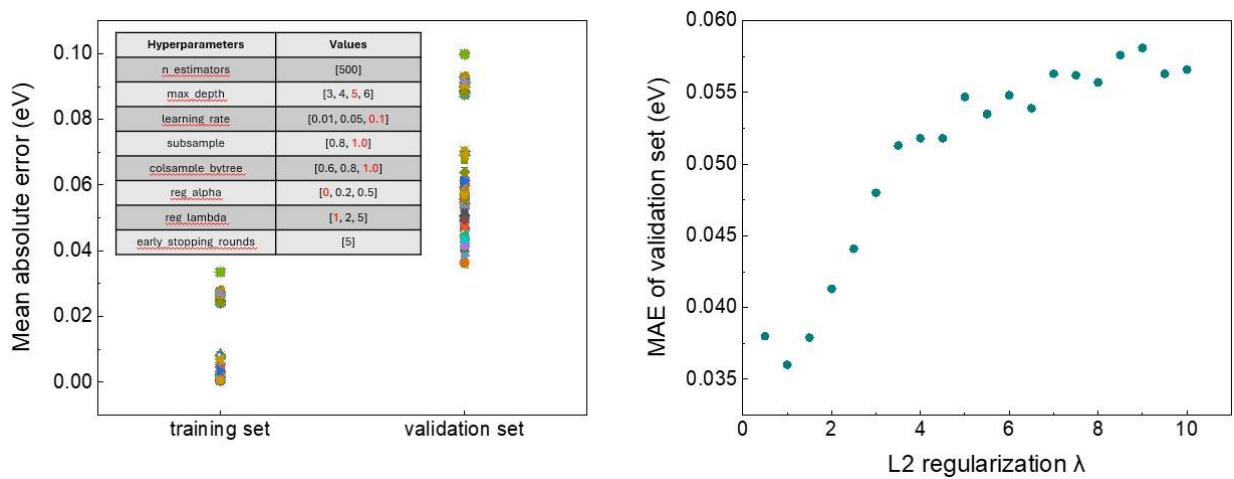

Figure S13. XGBoost model tuning with grid search and L2 regularization shows optimal λ = 1. However, generalization remains limited.

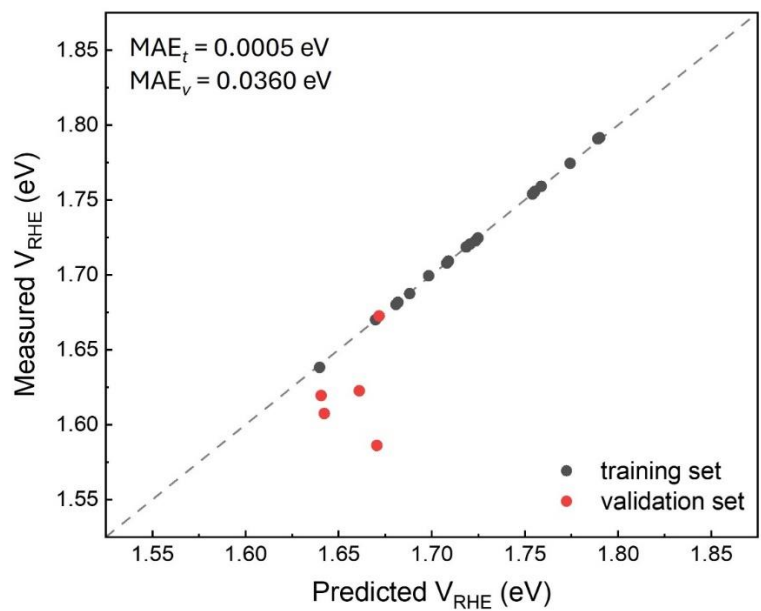

Figure S14. Best XGBoost model performance. Despite high flexibility, the model overfits the small dataset, with near-perfect training accuracy and weaker validation performance.

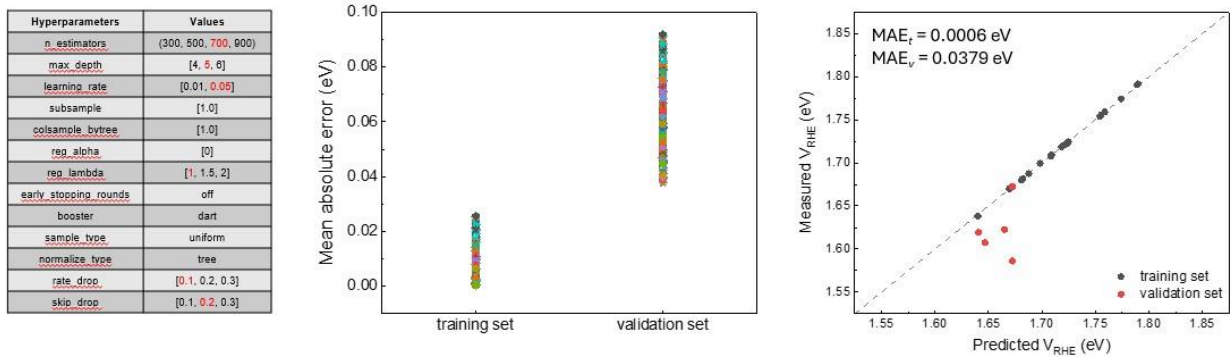

Figure S15. Performance of XGBoost with the DART booster. Hyperparameter search was conducted. Despite more aggressive drop-out regularization, overfitting remains evident in the validation MAE.

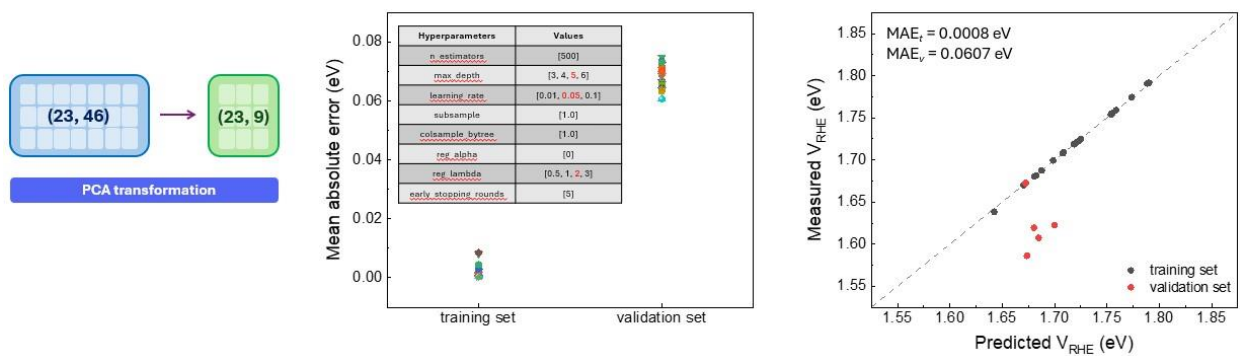

Figure S16. PCA reduces the feature set from 46 to 9 principal components. Even after dimensionality reduction, the XGBoost model shows poor generalization, with validation MAE remaining significantly higher than that of the neural network.

Table S1. The two best-performing formulas for each formula length group among 432 candidate formulas generated with the same dataset as reported by Weng *et al*. and their corresponding training MAEs (MAE$_t$).

| Formula length | Formula 1 | MAE$_t$ (eV) | Formula 2 | MAE$_t$ (eV) |
|---|---|---|---|---|
| 4 | $(\frac{\mu}{0.137})^{0.5}$ | 0.023 | $\frac{\mu^{0.5}}{0.37}$ | 0.023 |
| 5 | $2\mu + 0.906$ | 0.024 | $2\mu + 0.91$ | 0.024 |
| 6 | $(\chi_A + \frac{\mu}{0.214})^{0.5}$ | 0.021 | $\mu + (\chi_A + 0.681)^{0.5}$ | 0.021 |
| 7 | $\mu(\chi_A + 0.656) + t$ | 0.021 | $\mu(\chi_A + 0.655) + t$ | 0.021 |
| 8 | $(\chi_A + 0.279)^{0.5} + \mu R_A$ | 0.019 | $\chi_A^{0.5} + \mu(R_A + 0.311)$ | 0.019 |
| 9 | $\mu(\chi_A + \frac{\mu}{0.612}) + t$ | 0.018 | $(Q_A^{0.5} + \frac{t + R_A}{\chi_B})^{0.5}$ | 0.019 |
| 10 | $(\frac{R_A}{N_d} + \mu + \chi_A)^{0.5} + \mu$ | 0.017 | $\mu + (R_A\chi_A + \frac{0.487}{\chi_B})^{0.5}$ | 0.018 |
| 11 | $\mu(R_A\chi_A + 0.293) + t^{0.5^{0.5}}$ | 0.018 | $\mu + (Q_A^{0.5^{0.5}} + (\frac{\chi_A}{N_d})^{0.5})^{0.5}$ | 0.018 |
| 12 | $((\mu + R_A - \chi_B^{0.5})^{0.5} + \chi_A)^{0.5} + \mu$ | 0.018 | $(R_A + Q_A(R_A - t)^{0.5})^{0.5^{0.5}} + \mu$ | 0.019 |

Table S2. Formula candidates marked as points A-G in MAE distribution plots and their corresponding training and validation MAEs.

| Point | Formula | MAE (eV) (train, val) | Point | Formula | MAE (eV) (train, val) |
|---|---|---|---|---|---|
| A | $\frac{2\mu}{t} + 0.922$ | (0.0227, 0.0275) | G | $\frac{1.754}{t}$ | (0.0266, 0.0244) |
| B | $\frac{2\mu}{t} + 0.851^{0.5}$ | (0.0227, 0.0276) | H | $(\frac{\mu}{R_A})^{0.5} + \frac{\mu}{t} + 0.775$ | (0.0228, 0.0208) |
| C | $\frac{R_A^2}{R_A * t - \frac{\mu}{R_A}}$ | (0.0224, 0.0299) | I | $t + \frac{\frac{\mu}{t}}{\log R_A^{0.5}}$ | (0.0227, 0.0193) |
| D | $\log(R_A * t + \frac{\mu}{R_A} + \frac{\frac{\mu}{R_A}}{0.076})$ | (0.0224, 0.0274) | J | $t^{0.5^{0.5^{0.5}}} + \frac{\frac{\mu}{R_A}}{0.409}$ | (0.0236, 0.0206) |
| E | $(\frac{\frac{\mu}{R_A} + t^{0.5}}{0.736 - \frac{\mu}{R_A}})^{0.5}$ | (0.0227, 0.0205) | K | $(\frac{\frac{\mu}{R_A} + t^{0.5}}{0.736 - \frac{\mu}{R_A}})^{0.5}$ | (0.0227, 0.0205) |
| F | $(R_A * t)^{0.5^{0.5}} + (\frac{\frac{\mu}{t}}{\frac{\mu}{t} - R_A * t})^{0.5}$ | (0.0225, 0.0299) | L | $(0.398 + \frac{\mu}{t})^{0.5} + (\frac{\frac{\mu}{R_A}}{0.189^{0.5}})^{0.5}$ | (0.0229, 0.0209) |

Table S3. Refined symbolic regression formulas using μ/t, μ/R$_A$, and LUMO energy only. Two formulas among 432 candidate formulas include all three features. The top-performing one achieves validation MAE of 0.0206 eV with strong interpretability.

| Formula | Formula length | MAE$_t$ (eV) | MAE$_v$ (eV) |
|---|---|---|---|
| $\frac{\mu}{t} + \frac{\mu}{R_A} + E_{LUMO} + (-E_{LUMO})^{0.5} + 0.78$ | 10 | 0.0221 | 0.0206 |
| $(\frac{\mu}{t} + \frac{\mu}{R_A})^{0.5} + (E_{LUMO} + 0.91)^{0.5^{0.5}}$ | 10 | 0.0240 | 0.0283 |